\documentclass[prb,aps,twocolumn,showkeys,showpacs,superscriptaddress]{revtex4}
\usepackage{graphicx,epsfig}
\usepackage{bm}  
\usepackage{dcolumn}  
\def\bea{\begin{eqnarray}}
\def\eea{\end{eqnarray}}
\def\be{\begin{equation}}
\def\ee{\end{equation}}

\begin{document}
\title{Systematic Study of Electron Localization in an Amorphous 
Semiconductor}

\author{Raymond~Atta-Fynn}
\affiliation{Department of Physics and Astronomy,
Ohio University, Athens, Ohio 45701-2979, U.S.A.} 
\email{drabold@helios.phy.ohiou.edu}

\author{Parthapratim~Biswas}
\affiliation{Department of Physics and Astronomy,
Ohio University, Athens, Ohio 45701-2979, U.S.A.} 

\author{Pablo~Ordej{\'o}n}
\affiliation{Department of Physics and Astronomy,
Ohio University, Athens, Ohio 45701-2979, U.S.A.} 
\affiliation{Institut de Ci\`encia de Materials de Barcelona (CSIC),
Campus UAB, 08193 Bellaterra, Spain}

\author{D. A.~Drabold}
\affiliation{Department of Physics and Astronomy,
Ohio University, Athens, Ohio 45701-2979, U.S.A.}


\begin{abstract}

We investigate the electronic structure of gap and band tail
states in amorphous silicon. Starting with two 216-atom models 
of amorphous silicon with defect concentration close to the 
experiments, we systematically study the dependence of electron 
localization on basis set, density functional and spin polarization 
using the first principles density functional code {\sc Siesta}. 
We briefly compare
three different schemes for characterizing localization: 
information entropy,
inverse participation ratio and spatial variance.
Our results show that to accurately describe defected structures 
within self 
consistent density functional theory, a rich basis set is necessary. 
Our study revealed that the localization of the wave 
function associated with the defect states decreases with larger basis
sets and there is some enhancement of localization from GGA relative to LDA.
Spin localization results obtained via LSDA calculations, 
are in reasonable agreement with experiment and with previous 
LSDA calculations on \emph{a}-Si:H models. 

\end{abstract}

\maketitle
\section{INTRODUCTION}

Amorphous semiconductors represent a large and important area in materials 
science, with interest both from the technological and fundamental 
point of view.  Coordination defects in \emph{a}-Si are of key importance 
to bulk and transport properties. Understanding the nature of defects in 
\emph{a}-Si can help unlock the mystery behind  phenomena 
like the Staebler-Wronski effect~\cite{Elliot} and help establish the link 
between localization 
of defect states and large electron-phonon coupling. Early theoretical work 
on defect states in \emph{a}-Si and \emph{a}-Si:H was based on tight-binding 
methods~\cite{Biswas,Allan,Fedders,Holender,Kneif,Min}. For example, 
Biswas~\emph{et al.}~\cite{Biswas} and Fedders and Carlsson~\cite{Fedders} 
investigated the electronic structure of dangling and floating bonds 
in \emph{a}-Si. 
They showed that gap defect states associated with dangling 
bonds are strongly localized on the central atom of the dangling 
bond~\cite{Biswas}. 
More recently, density functional calculations of 
dangling bond states using 
the local density approximation (LDA)~\cite{Kohn1} 
have been performed by Fedders 
and Drabold~\cite{Drabold1}. They reported a wave function 
localization of 10-15\% 
on the central atom in supercell models with one defect and far less on 
supercell models with many defects due to defect band formation. 
This finding was at 
variance with ESR experiments, which showed that over 50\% of spin density 
of the gap state is located on the central atom of the 
dangling bond~\cite{Biegelsen,Umeda}. 
However, recent calculations by Fedders \emph{et al.},~\cite{Drabold2} 
using the local spin density approximation (LSDA) have shown that a large spin 
localization of a defect state does not necessarily imply
the existence of a corresponding wave function localization. 
   
Density functional theory (DFT)~\cite{Kohn2} has enjoyed enormous success in
describing the ground state properties 
and defects for a wide range of materials,
and in particular, \emph{a}-Si. 
Nonetheless, the one-particle Kohn-Sham energies in the theory have no formal 
justification as quasiparticle energies. However from an empirical
point of view, Hybertsen and Louie have shown, 
using GW calculations,~\cite{Louie1} 
that for states close to the fundamental 
band edges of bulk semiconductors and in 
particular Si, there is a 99.9\% overlap 
of the quasiparticle wave function 
with the corresponding Kohn-Sham orbital 
(GW calculations provide post Hartree 
self-energy corrections to DFT/LDA). 
This provides some rationale for interpreting 
the Kohn-Sham orbitals as quasiparticle states.

Within the density functional framework, there is a general problem for 
the accurate representation of localized midgap and band tail states in 
amorphous semiconductors. The reliability of density 
functionals to correctly 
reveal the localized behavior of electronic 
states with respect to its wave function 
and spin is very important. In particular, 
the generalized gradient approximation (GGA) and 
LSDA results for electron localization are sometimes quite different. 
For example, recent density functional and GW studies of the metal-insulator 
transition of bcc hydrogen showed that eigenfunctions obtained GGA 
are more localized~\cite{Louie2,Louie3} 
and closer to quasiparticle energies and states 
than LSDA. Also, it was observed that GGA bandgap was systematically larger 
than LSDA gap\cite{Louie2}. 
  
In this paper, we systematically examine 
the localization of band tail and gap states 
and its dependence on basis sets, density 
functionals and spin polarization for three 
defected models: two 216-atom supercell models 
of amorphous Si and a 218-atom supercell 
model of crystalline Si:H with a vacancy. 
The crystalline model will serve as 
a benchmark to compare the nature of a 
dangling bond defect state in an amorphous 
environment  with that of the crystalline 
phase. We compare the localization of gap and tail 
states within the LDA, LSDA and GGA for
 frozen lattices (unrelaxed samples) as well 
as for samples fully relaxed for a given Hamiltonian. 
Our motivation for performing the frozen 
lattice calculation is to systematically 
investigate the sole effect of different basis sets and 
density functionals on the localization of states for a fixed configuration. 
We study the relaxation effects to see the dependence of the local geometry of 
the defect sites on the different basis orbitals and density 
functionals and the 
behavior of localized defect states in a 
relaxed environment compared to the frozen one. 
We computed spin and wave function localization for defect states,
to determine the correlation between spin density and charge density.

\section{METHODOLOGY}
\subsection{Models and calculations}
Two of the models used here were 
generated by Barkema and Mousseau~\cite{Moose} 
using an improved version of the Wooten, 
Winer and Weaire (WWW) algorithm~\cite{Wooten}. The 
details of the construction was reported in Ref. 19 and we just give a brief 
recap here. Two independent 216-atom models of \emph{a}-Si, each 
with two dangling bonds, were generated.
Atoms are first packed randomly in a box, 
at crystalline density, with the single 
constraint that no two atoms be closer than 2.3 {\AA}. A connectivity table was 
then set up by constructing a loop that passes exactly twice through each atom. 
To create a dangling bond, a ghost bond 
(bond with zero energy) is placed between 
two atoms. The two atoms are chosen in such a way that they are quite close in 
one model and reasonably far in another model. The network is 
then relaxed using 
a fixed list of neighbors and a Keating potential. These steps ensure that the 
initial state has no trace of crystallinity. The models are then 
relaxed using a 
series of WWW moves, using the accelerated algorithm 
discussed in Ref. 5. The conformations 
are finally relaxed at zero pressure 
with a Keating potential. In this work we 
refer these two 216-atom supercell models 
of \emph{a}-Si as CLOSE and FAR. In model 
CLOSE the atoms with the dangling bonds are separated 
by a distance of 4.6 {\AA}, whereas in the model FAR they are 7.7 {\AA} apart.
The third model, which we will refer to as 
\emph{c}-Si:H, is generated as follows: 
starting with a 216-atom cell of silicon in the diamond structure, 
an atom is removed resulting in the creation of a vacancy with
four dangling bonds. 
Three of them were terminated by placing an H atom at about 1.5 {\AA} from 
the corresponding threefold-coordinated Si atom. This results 
in one isolated dangling bond. Throughout this work,
we will use a cut-off radius $\rm R_{\rm Si-\rm Si}$ = 2.6 {\AA} to 
define Si-Si coordination.

Our DFT calculations were performed within the  LDA (with and 
without spin polarization) and the GGA using the code 
{\sc Siesta}\cite{Siesta1,Siesta2,Siesta3}. We used the 
parameterization of Perdew and Zunger~\cite{Perdew1} 
for the exchange-correlation 
functional in all LDA calculations  and that of Perdew, Burke and 
Ernzerhof~\cite{Perdew2} for the exchange-correlation functional in all 
GGA calculations. Norm conserving 
Troullier-Martins pseudopotentials ~\cite{Troullier} 
factorized in the Kleinman-Bylander 
form~\cite{Kleinman} were used to remove 
core electrons. To describe the valence electrons, we used 
atomic orbitals basis set consisting of finite-range numerical pseudoatomic 
wave functions of the form proposed by Sankey and Niklewski~\cite{Sankey}. We 
employ single-$\zeta$ (SZ) and double-$\zeta$  
with polarization functions (DZP) 
basis sets on all the atoms\cite{basis}.  
We solved the self consistent Kohn-Sham equations by
direct diagonalization of the Hamiltonian.
The $\Gamma$
point was used to sample the Brillouin zone in all calculations.

\subsection{Quantifying the degree of localization}

To characterize the spatial extent of an 
electronic state we use three measures of localization. 
Two of them depend on the Mulliken 
(point) charge~\cite{Szabo} $\emph{q}_{i}(\rm E)$ 
residing at an atomic site $i$ 
for an eigenstate with energy eigenvalue $\rm E$. The charges are normalized 
(${\sum_{i=1}^{\rm N}{\emph{q}_{i}(\rm E)}} = 1$, 
where $\rm N$ is the total number of atoms in the supercell). 
The third is a measure of the spread of the
wave function in real space, as the 
the second moment of the radial displacement 
about its center (variance).
We briefly describe below each measure and its interpretation.

The first quantity, the conventional inverse participation ratio (IPR)
\be 
\rm I(\rm E) = {\sum_{i=1}^{\rm N}({\emph{q}_{i}(\rm E)})^{2}},
\ee
is a measure of the inverse of the number of 
sites involved in the state with energy $\rm E$. 
For a uniformly extended state, the Mulliken 
charge contribution per site is uniform and equals 
${ \frac{1}{\rm N} }$ and so $\rm I(\rm E) = { \frac{1}{\rm N} }$. 
For an ideally localized 
state, only one atomic site contributes all the 
charge and so  $\rm I(\rm E) = 1$. This 
implies that high values of $\rm I(\rm E)$ correspond to localized states and 
low values of correspond to delocalized states. 
Despite its ubiquity, the IPR is an \emph{ad hoc} measure 
of localization.
The second quantity has been used by Lewis 
\emph{et.~al}~\cite{Sankey2,Sankey3} to measure
the localization of DNA electron states. 
It utilizes the information entropy~\cite{Shannon,Jaynes} 
$S(\rm E)$ of an eigenstate with energy $\rm E$ and is given by
\be 
\rm W(\rm E) = \exp(S(\rm E))
\ee
where 
$S(\rm E) = -{\sum_{i=1}^{\rm N}
{\emph{q}_{i}(\rm E)}ln\,{\emph{q}_{i}(\rm E)}}$. 
In information theory, the entropy is a 
measure of disorder or randomness in the \{$\emph{q}_i$\}.
As such, it has more \emph{a priori} justification 
for use as a measure of localization. This is 
especially true if we interpret  localization as 
a {\it departure from randomness}\cite{ed}.
We interpret $\rm W(\rm E)$ as an estimate of the \emph{number 
of accessible atoms} for a given state  
or equivalently, the number of atoms a particular electronic state can reach.
It follows that $\rm W(\rm E)$ ranges from the total number of atoms for a 
uniformly extended state (that is, all the atoms are accessible) to unity
for an ideally localized state (that is, 
only one atom is accessible). For purposes of 
discussions and comparisons to IPR, we use 
$\frac{1}{\rm W(\rm E)}$ for our work.

The third quantity we use to determine the localization
of the wave functions is a measure of the spread in
real space, as defined in the context of 
localized wave functions in molecules~\cite{boys}
and maximally localized Wannier functions in solids.~\cite{mlwf,wannier}
This definition assumes that the wave function
has a well defined center, and that the average
of the position operator can be defined.
This is not the case for our eigenstates,
which correspond to the $\Gamma$ point 
and therefore are periodic in the supercell.
However, since we are interested in states which
are localized (exponentially) within the supercell,
we can still use the definition of the spread,
by integrating over the cubic supercell volume
centered at the defect site:
\be
\sigma^{2} =  [{\langle {{\bf r}^2} \rangle} - 
{\langle {\bf r} \rangle}^2]. 
\ee
We also compute the amount of the norm of the wave function
that is around the defect center, by integrating
over a sphere $\Omega$ centered at the defect and which has
the same diameter as the lattice constant
(so it is the largest sphere that fits into the supercell 
volume).
Clearly, the defect states will not be completely
localized within the sphere, and the
norm of the wave function inside the integration volume,
$q­_\Omega$,
will be smaller than one. This is also a measure of
the amount of wave function which is outside
the integration sphere: for an extended state,
the amount of charge will be close to one half,
since the sphere encloses approximately half the
supercell volume. For localized states, we typically
capture more than 80\% of the wave function norm.
We will characterize the localization by using both 
the spread $\sigma^2$ (larger values correspond to 
less localization) and by the norm $q­_\Omega$.

\section{Results}

\subsection{Frozen lattice calculations}

\subsubsection {Wave function localization}

\begin{figure}[ht]
\includegraphics[width=\linewidth]{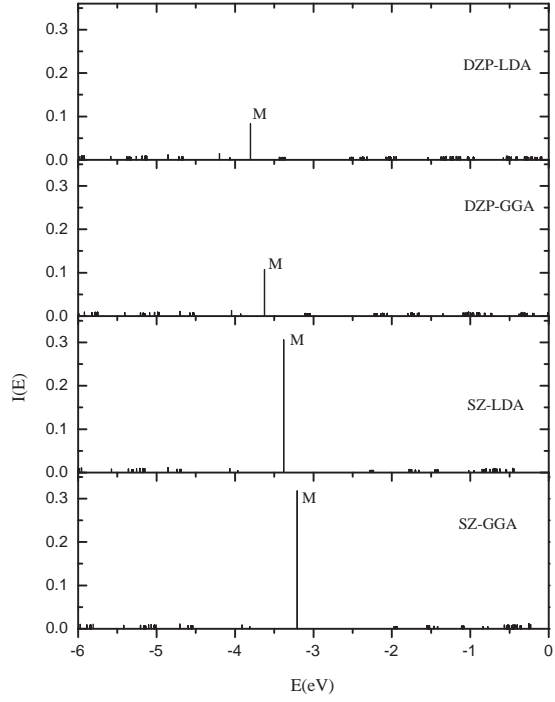}
\caption{\label{ipr-cSi}
IPR for the model $c$-Si:H computed using frozen coordinates. The
highly localized midgap state (labeled M) sits on the 
central atom of the dangling bond. For the SZ basis functions 
the charge localization on the central atom 
within the LDA and GGA are respectively $0.54e$ and $0.57e$. For 
the DZP basis sets, the charge localization on 
the atom reduce to $0.27e$ and $0.31e$ respectively 
within the LDA and GGA.}
\end{figure}

\begin{figure}[ht]
\includegraphics[width=\linewidth]{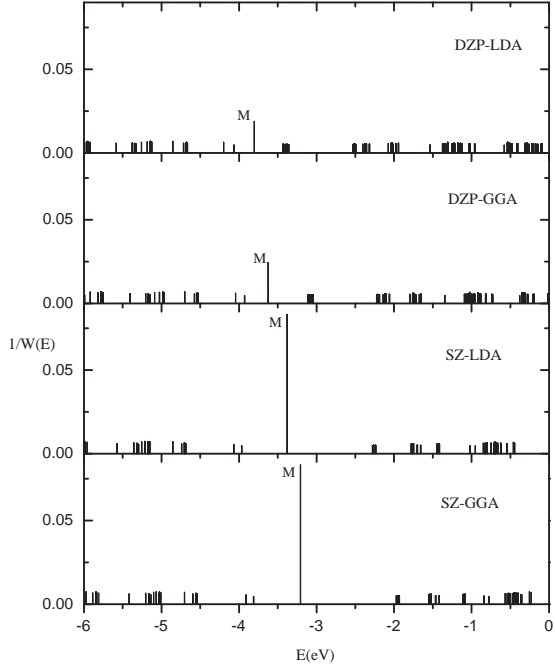}
\caption{\label{w-cSi}Localization of states 
(as measured using $\frac{1}{\rm W(\rm E)}$)
for the model \emph{c}-Si:H using the frozen 
lattice. 
The onlylocalized midgap state (labeled M) sits on the central atom of the 
dangling bond.}
\end{figure}

In order to understand the effect of basis set and
density functional on the localization of defects states,
we first compute the electronic structure of the three
models with fixed geometry, using the different basis sets
and density functionals.
We begin our 
discussion with simplest of the three models, 
{\it c}-Si:H.
The position and localization for individual states are
reported in Figs.~\ref{ipr-cSi} and \ref{w-cSi}, where the IPR
and $\frac{1}{\rm W(\rm E)}$, respectively, are used to measure
localization. Each spike indicates 
an energy eigenvalue.
With both measures of localization, and for all basis sets and density
functionals considered, we observe a highly localized 
state near at Fermi level,
with all the other states being extended.
This state is centered on the dangling bond atom,
which contributes with a Mulliken charge
of $0.54e$ and 
$0.57e$ in the SZ-LDA and SZ-GGA cases
respectively, and $0.27e$ and $0.31e$ in the DZP-LDA and DZP-GGA cases
respectively. The rest of the state is mainly localized
in the neighbor atoms to the dangling bond.
Both the IPR and 
$\frac{1}{\rm W(\rm E)}$  provide the same
qualitative picture of the localized state, and
its evolution with basis set and density functional.
In particular, localization decreases strongly when we go from the SZ basis
to a more complete DZP basis. It also increases but only
by a small amount when using GGA instead of LDA. 
These are general trends which we will also observe
with the other models, as we will see below.
We note that the decrease in the measure of localization is
not proportional to the decrease in the Mulliken charges
at the defect state, as both definitions of localization
are nonlinear. While the charge at the dangling bond site
is reduced by a factor of two when moving from SZ
to DZP, the localization measure decreases by roughly a factor of three, 
both for IPR and $\frac{1}{\rm W(\rm E)}$.

\begin{figure}[htbp]
\includegraphics[width=\linewidth]{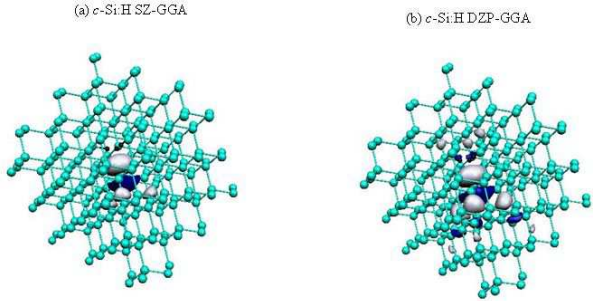}
\caption{\label{rho-cSi} (Color online) Isosurface plots 
for localized wave functions corresponding to a \emph{c}-Si:H 
dangling bond defect state. The wave functions were generated with 
the same cut-off. Each surface is labeled 
according to the basis set and functional used. The surface 
is confined to a small region in space in the simple SZ case. 
For the DZP basis we see a localized dangling bond orbital 
with pieces of the surface in the vacancy and other neighboring 
sites. H atoms are colored black.}
\end{figure}

Unlike $\rm I(\rm E)$ and $\rm W(\rm E)$ that 
are point estimators of the localization (in the
sense that they use only the Mulliken charge at each atomic site), 
$\sigma^2$ is a more physical representation.
The results obtained for $\sigma^2$ for 
\emph{c}-Si:H with the lattice frozen are presented 
in Table~\ref{tab1}. The spread for the
localized state $\rm M$  is simple to 
compute since it is unimodal (peaked at
only one site and therefore having a well defined center). 
From $\sigma^2$, we also see monotonic increase in the
spread and a decrease in the total charge in the 
localization volume as the basis sets are increased 
(see Table~\ref{tab1}). The GGA states
show a slightly smaller spread than the corresponding LDA states. 
In Fig.~\ref{rho-cSi} we show snapshots of the 
isosurface of the wave function for the localized midgap state
of \emph{c}-Si:H  within two approximations. 
We see a dangling bond orbital confined
to a small region in space in the SZ case 
implying that the state is well localized. 
In the DZP case, we see pieces of the  
surface in the vacancy and other neighboring atoms besides the
dangling bond orbital making it less localized compared to the SZ case.

\begin{figure}[htbp]
\includegraphics[width=\linewidth]{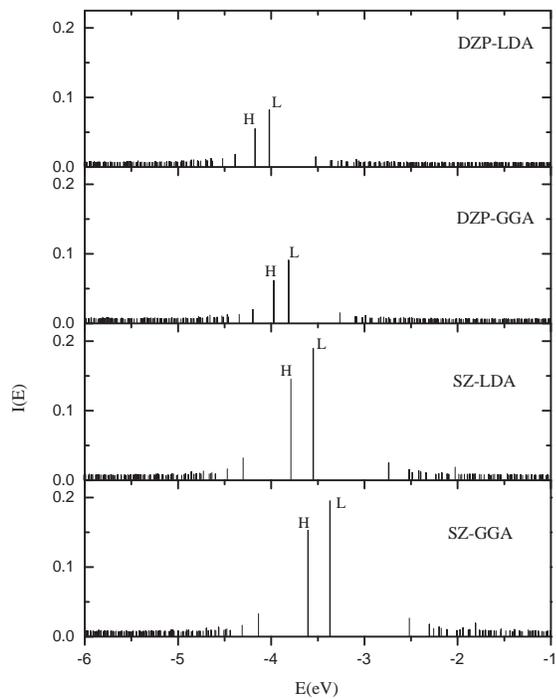}
\caption{\label{ipr-close}IPR for the model CLOSE 
using frozen coordinates. The
two highly localized midgap states sit on the 
central atoms of the two dangling bonds.
The state labeled H is the highest occupied 
molecular orbital (HOMO) and the state labeled 
L is the lowest unoccupied molecular orbital (LUMO).}
\end{figure}

\begin{figure}[htbp]
\includegraphics[width=\linewidth]{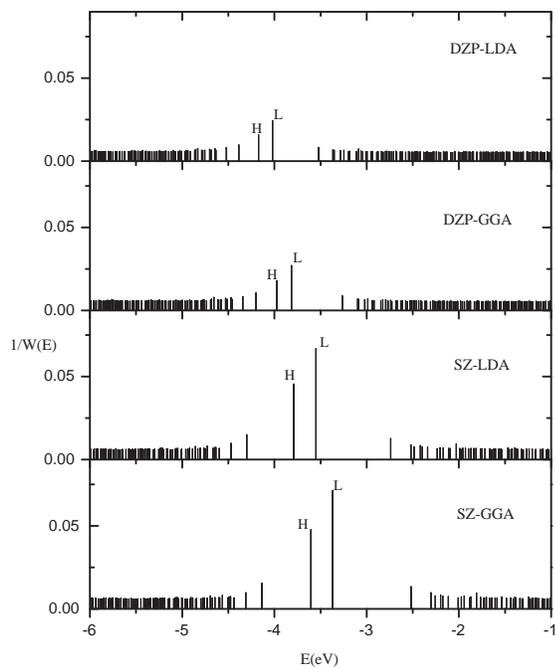}
\caption{
\label{w-close}
$\frac{1}{\rm W(\rm E)}$  for the model CLOSE with the atoms frozen.}
\end{figure}

Next, we analyze the localized nature 
of the states for the amorphous model CLOSE. 
In this system, we expect to see two localized states in the
gap, corresponding to the two dangling bonds
present in the structure.
Indeed, we see two highly localized gap states H and L in 
the IPR shown in Fig.~\ref{ipr-close}, with an energy splitting which 
is just over a tenth of an eV. The state H is the highest 
occupied molecular orbital (HOMO) and L is 
the lowest unoccupied molecular orbital (LUMO).
Due to the small distance between the two dangling bonds,
the localized H and L states are bonding and antibonding
combinations of the dangling bond states, and therefore
both H and L have nearly equal weights on the two defect
sites.
In the SZ case, the total Mulliken charge contributions 
from the two dangling
bond sites for each of the two localized wave 
functions range between $0.52e$ and $0.64e$.
The charge concentrations
drop to the range $0.29e$-$0.40e$ in the DZP case. 
In Fig.~\ref{w-close}, we show the results for $\frac{1}{\rm W(\rm E)}$. 
We see a sharp drop in the number of atoms a 
given eigenstate can reach as the 
energy changes from the edge of either the valence or conduction band
into the gap. 
We again see that  all the features of $\rm I(\rm E)$ are
reproduced in $\frac{1}{\rm W(\rm E)}$. For each localized state, 
both measures decrease by approximately a factor of 
2 when the basis sets are increased from SZ to DZP. 

\begin{figure}[htbp]
\includegraphics[width=\linewidth]{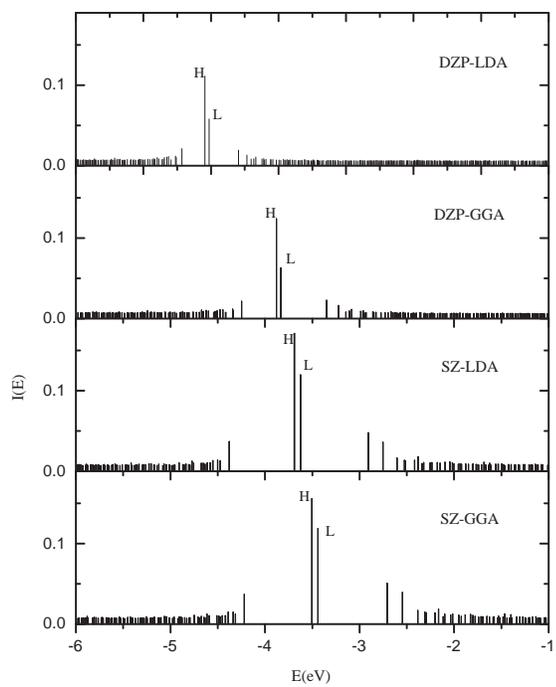}
\caption{\label{ipr-far}IPR for the model FAR using frozen coordinates. The
two highly localized midgap states (H for HOMO and L for LUMO)
are nearly degenerate.}
\end{figure}

In Fig.~\ref{ipr-far}, we show the IPR for the model FAR.
As in the CLOSE case, both the HOMO and LUMO states
are localized. Now, however, since the distance
between the two dangling bond sites is larger, the HOMO-LUMO
splitting is much smaller, only $\sim 6$ meV. 
The HOMO has now most of its weight on one of the
dangling bonds, whereas the LUMO is mostly localized
in the other.
The trend in localization of the gap 
states is similar to the the trend observed in the other two two models,
decreasing strongly with more complete basis sets.

Since the wave functions associated with the gap states in the unrelaxed
CLOSE and FAR models do not have a single center, but
are peaked at the two dangling bonds, we will not analyze the
localization by means of the spread in these cases.
However, as we show in next section, relaxation leads
in some cases to localization of the wave functions around
one of the dangling bonds, and this will allow us to use
the spread in such cases to quantify localization.

Our frozen lattice calculations consistently show 
that increasing the basis set decreases 
the localization. Although this is not unexpected, the huge 
decrease in the localization of the wave function as 
the size of the basis functions 
increase from SZ to DZP is quite surprising. The fact that 
the results are consistent in both the amorphous 
and crystalline system makes it even more
interesting and general.  A plausible explanation
for this effect is that the energy gap is clearly
reduced as the basis set is more complete.
The localized states are then closer to the 
band edges, and therefore are more able to mix with the
extended bulk states, becoming more delocalized.
Obviously, the delocalization will not proceed
indefinitely upon improvement of the basis set,
but will converge as the basis approaches
completeness.

\subsubsection{Spin localization versus wave function localization}

Fedders and co-workers~\cite{Drabold2} have shown that, in order
to correlate the degree of localization from dangling bond
states with ESR experiments, it is not enough to look
at the wave functions, but to the net spin polarization
near the dangling bond. The reason is that the spin
density has also contributions from electronic states other than the
localized defect wave function, which contribute to make
the spin polarization more localized than the specific
localized state wave function.
In order to confirm this result (obtained by Fedders {\it et al.}
on cells of a-Si:H) in our structural models, we performed 
calculations allowing for spin polarization in our frozen
lattice models, using the DZP basis set.
Except for the $c$-Si:H case, where there is one unpair electron
that yields a net spin polarization, we were not able
to find a spin polarized solution for any of the
amorphous cells. The reason is the existence of
two interacting dangling bonds, which favors the formation
of a spin singlet with two electrons paired.
In order to force the appearance of a spin moment
in our models, we introduce an unpaired spin by removing
a single electron from the system.

In the model \emph{c}-Si:H with, we find a contribution of almost 50\%
to net spin by the central dangling bond and its neighbors 
(the central atom alone contributing 38\%).
However, the Mulliken charge contribution to the wave function
of the corresponding localized state from the defect site 
is only $0.29 e$. The hydrogen-terminated dangling bond
sites also contribute 
about 10\% of the net spin. The remainder
is somewhat distributed uniformly at the other sites.
In CLOSE, about 57\%
of the net spin polarization was due the one dangling bond and its three
neighbors. The other dangling bond contributed only 6\% to the net spin
with essentially zero contribution coming from the neighbors.
In FAR, about 54\% of the net spin
localization sits on the isolated dangling bond and its nearest neighbors.
Our results are in good agreement with
the previous LSDA calculations by Fedders {\it et al.}\cite{Drabold2},
and in reasonable agreement with the experiment~\cite{Biegelsen,Umeda}.

Our results confirm that, for a dangling bond defect state, there is a
rather large difference between spin localization
and wave function localization. In particular, the degree
of spin localization is greater than that of the wave function localization
at the dangling bond site. To our
knowledge, no experimental methods exist for 
measuring the extent wave function
localization on the dangling bond orbital as opposed to spin

\subsection{Relaxation effects}

\subsubsection{Geometry of defect sites, density functionals and basis sets}
  
In this section we discuss the geometry 
around the defect sites before and after relaxation. The details
of the local geometry are very important in determining the local 
electronic structure and the strain around the defect site, 
and here we study the dependence with varying basis 
and density functional. We relax all the models 
using a conjugate gradient optimization until the
largest atomic force is smaller than 0.04 ${\rm eV}$/{\AA}.

Relaxation effects for the simple dangling bond defect in
\emph{c}-Si:H are relatively small. There is no major
rearrangements in bonding, but only a relaxation of
the surroundings of the vacancy site. 

In the unrelaxed CLOSE model, the dangling bonds were 
originally separated  by a distance of 4.6 {\AA}. After SZ basis 
relaxations, both with LDA and GGA, 
the defect sites come closer, to a distance of about
2.6 {\AA}, to form a highly strained bond. 
For the more complete DZP basis set, the two defect
sites also approach each other, but they continue
being undercoordinated, so the two distinct dangling bonds
remain present.

In the case of the FAR model, 
one of the two well separated dangling bonds forms a strained bond 
with a neighboring atom after a SZ-LDA relaxation. 
The dangling bond is therefore 
terminated, and a floating bond is introduced. 
The other dangling bond remains present. 
For the SZ-GGA relaxation, the two dangling bonds 
still continue to exist, but the one that 
was terminated in the SZ-LDA case also approaches a
neighbor and tries to form a bond.
After the DZP 
relaxations, the dangling bonds still exist both in LDA and GGA.  

Our results indicate that SZ basis tends to favor 
tetrahedral bond formation whereas 
DZP qualitatively preserves the original
structure with the dangling bonds present. 
Also, the SZ tends to favor the transformation
of dangling bonds into floating bonds. The results therefore 
suggest that the richer DZP basis set is necessary for an 
accurate description of the geometry of 
both isolated and clustered dangling 
bonds in amorphous silicon. 
The SZ basis is not flexible enough to provide
sufficient freedom to describe the different
shape of the wave function at the dangling
bonds compared to covalent $sp^3$ bonds (for which the
basis is ideally suited). Therefore, it tends to
favor the disappearance of the dangling
bonds through annihilation with other dangling
bonds or formation of floating bonds with other
already fourfold coordinated atoms.

\subsubsection{Localization, density functionals and basis sets} 

We now consider in  detailed the trends in the 
localization behavior of the electronic states in the gap 
after full structural relaxations. 
We also examine the role the defect site in a relaxed 
environment plays in the localization of gap states. 
We study the localization using the IPR for 
each of the three fully relaxed 
models within the different approximations.

\begin{figure}[htbp]
\includegraphics[width=\linewidth]{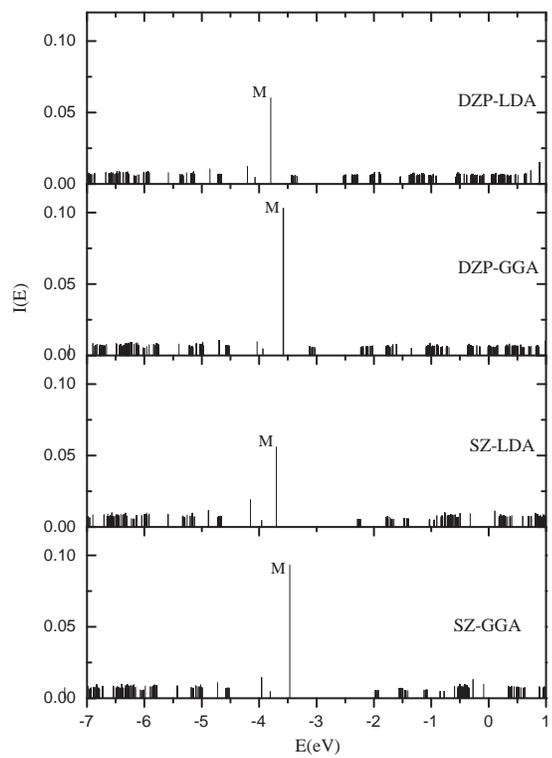}
\caption{\label{ipr-cSi-rel}IPR for the 
fully relaxed model \emph{c}-Si:H.}
\end{figure}

We first studied 
the simple dangling bond defect in
\emph{c}-Si:H. 
Fig.~\ref{ipr-cSi-rel} shows that, for the relaxed
structures, the localization
behavior of the midgap state is density
functional dependent but rather basis set independent,
contrary to the results obtained in the frozen 
lattice calculations (Fig.~\ref{ipr-cSi}).
In other words, within
the same density functional approximation, SZ yields
a similar wave function localization as DZP
for the simple defect in the relaxed crystalline
environment. We also see that the GGA defect
state is more localized than the LDA defect state for a given basis
set. The IPR values obtained with DZP are,
nevertheless, almost unchanged upon relaxation, 
the difference between the unrelaxed and relaxed
geometries occurring mainly for the SZ basis.
The analysis from the  
real space spread, shown in Table~\ref{tab2},
confirms the results obtained 
via the IPR. We see again the similarity between 
SZ and DZP for a given functional, with the GGA 
states having a smaller spread than LDA.

\begin{figure}[htbp]
\includegraphics[width=\linewidth]{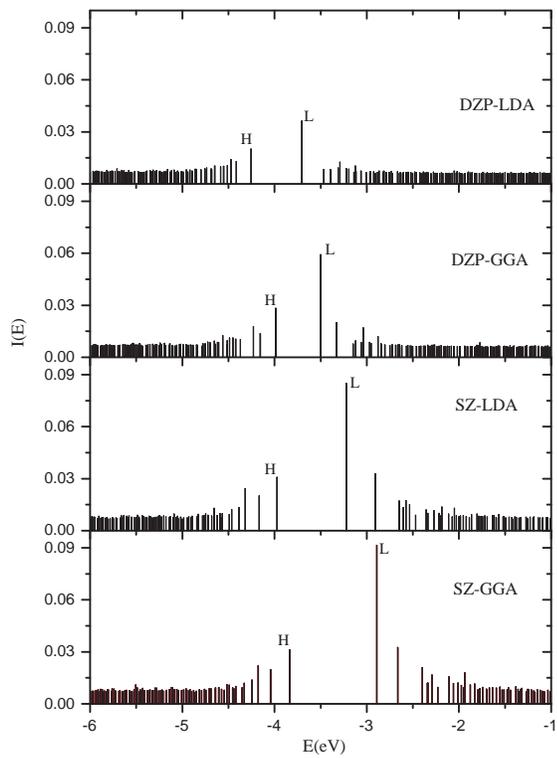}
\caption{\label{ipr-close-rel}IPR for the fully relaxed model 
CLOSE.}
\end{figure}

The IPR for the fully relaxed CLOSE is shown in Fig.~\ref{ipr-close-rel}.  
As discussed in the 
previous subsection, structural relaxation for this model is basis 
dependent. We first see that the splitting between the HOMO and
LUMO levels is now much larger than in the unrelaxes case.
This  is attributed to the fact that, in order to
minimize the
total energy, occupied defect states move down
in energy and closer into
the valence band edge, whereas the unoccupied states
do not affect the energy and thus can move
towards the conduction band edge. 
The second observation is that now the localization
of both HOMO and LUMO has decreased considerably compared
to the unrelaxed case. Again, this is a consequence of
the levels being closer to the band edges,
mixing more strongly with the delocalized
bulk states and therefore becoming less localized.
The effect is larger for the HOMO, which is the
one that adjusts its shape to optimize the total
energy.  A third observation is the appearance of
increasingly localized states in the band edges,
corresponding to bulk states which start becoming
localized and form the precursor of band tails.
This effect is originated from the strain field
imposed by the relaxation of the sites around
the defects. Therefore, the presence of defects
like dangling bonds in amorphous silicon also brings
the appearance of band tails of weakly localized states
due to the introduction of stress in the surroundings
of the defect. This supports results from previous
work~\cite{Drabold3} that there is not a one to 
one correspondence between spectral and geometric defects.  
The localization of the tail states
decays as one moves deeper
into the conduction and valence regions, as was 
previously observed by Dong and
Drabold~\cite{Dong} using a simple
orthogonal tight-binding Hamiltonian on a much larger 4096-atom
model of \emph{a}-Si. 

We now focus on the evolution of localization 
with basis set and density functional for the relaxed
CLOSE model.  As we observed with the $c$-Si:H case,
the HOMO level becomes less localized upon relaxation,
specially in the case of the SZ basis, for which the IPR
is reduced by more than a factor of two. 
The degree of HOMO localization predicted by the SZ and
DZP bases for the relaxed structure is therefore
very similar. For the LUMO, the difference between
SZ and DZP is still large, as in the unrelaxed case.
The high IPR values associated with the LUMO in 
the SZ cases are primarily 
due to strain (as a result of the bond between 
the two neighboring dangling bonds atoms).
The spread of the LUMO 
is reported in Table~\ref{tab3}. We again observe the common
trends in the spread:  GGA states show slightly less spread 
compared to LDA, and SZ basis set 
yield more localized states than DZP basis set. 

\begin{figure}[htbp]
\includegraphics[width=\linewidth]{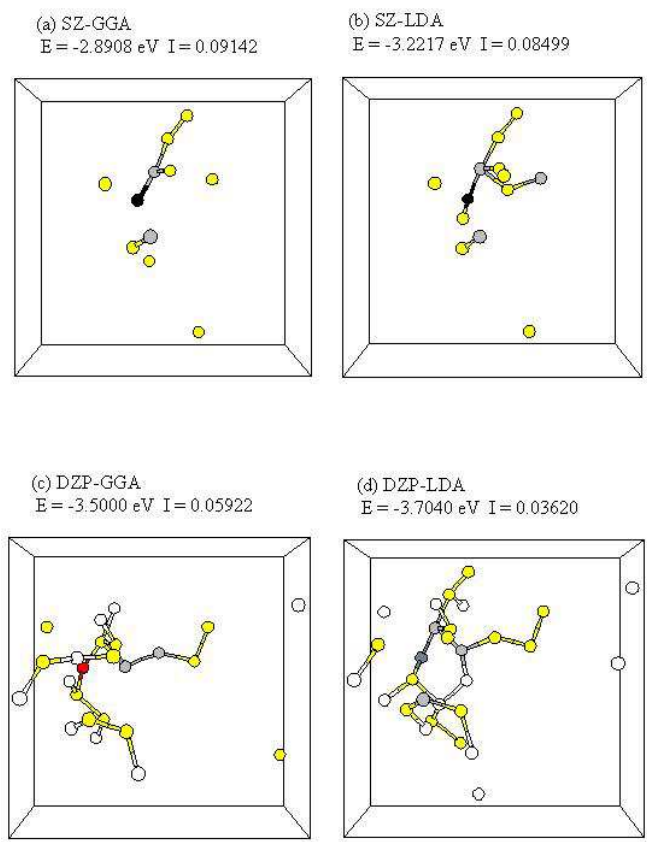}
\caption{\label{space-close} (Color online) Spatial character of 
localized eigenstates for the LUMO state for the relaxed CLOSE 
model. The energy  and its corresponding IPR localization are 
indicated in each picture. We use the following color code
to depict the fraction of the Mulliken charge $q$ for the 
localized state at each atomic site: black (q $>$ 0.25), 
red (0.15 $<$ q $<$ 0.25), slategrey (0.10 $<$ q $<$ 0.15), 
gray (0.05 $<$ q $<$ 0.10), yellow (0.01 $<$ q $<$ 0.05)
and white (q $<$ 0.01). Only 65\% of the total charge is 
shown.}
\end{figure}

In order to get a pictorial representation of
the localized states in this relaxed model and
the evolution with basis set and density functional,
we assign different 
colors to each site according to its
Mulliken charge contribution to the given eigenstate. We depict
this spatial feature by showing only 65\% of the 
total charge for the LUMO in Fig.~\ref{space-close}. 
We observe a small network connection of atoms for the localized
states in Figs.~\ref{space-close}(a) and \ref{space-close}(b) but the 
connectivity spreads out in a rather 1D fashion, mimicking a chain in
Figs.~\ref{space-close}(c) and \ref{space-close}(d). The
small size of our cell does not allow us to
immediately visualize a localized region containing completely
the cluster of atoms, which can be done in larger supercell containing
thousands of atoms~\cite{Dong}.

\begin{figure}[htbp]
\includegraphics[width=\linewidth]{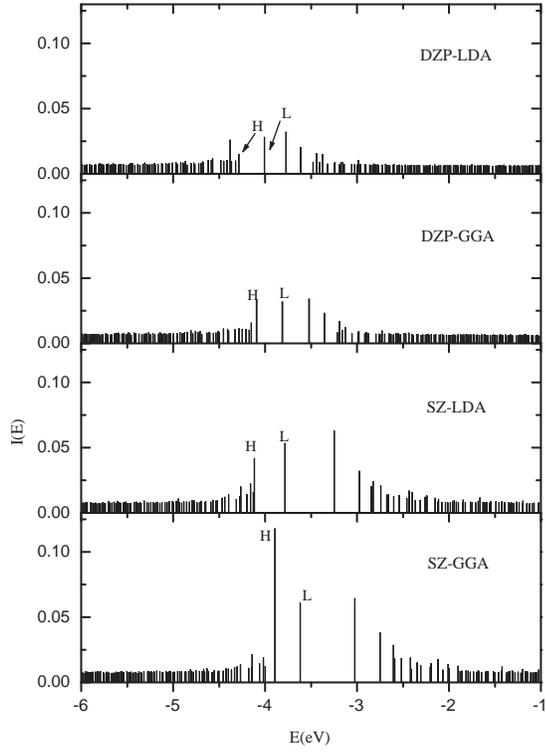}
\caption{\label{ipr-far-rel}IPR for the fully relaxed model FAR.}
\end{figure}

We now turn to the relaxed FAR model.
In Fig.~\ref{ipr-far-rel}, we have plotted  the IPR for the 
this model, for 
different density functionals and basis sets. 
Again, we see the same general features 
that we pointed out in the relaxed
CLOSE case. First, the splitting between HOMO
and LUMO is much larger than in the unrelaxed case.
Both states, and specially the HOMO, become
more delocalized upon relaxation, with the notable
exception of the SZ-GGA case, which yields a strongly
localized HOMO state. We also see the
formation of band tail states, and even the 
presence of strongly localized states in the
gap above the LUMO, due to strong relaxation
induced strain fields.
Again, in this model, we observe that
localization is stronger for GGA than with LDA,
and that the difference
in localization between SZ and DZP bases is
much reduced upon relaxation. 
The difference in localization for the SZ-GGA and
SZ-LDA cases can be explained by the fact 
that, as mentioned previously, SZ-LDA relaxation
results in the disappearance of a dangling bond and
the formation of a floating bond, which are known
to be less localized than dangling bond defects.

\begin{figure}[htbp]
\includegraphics[width=\linewidth]{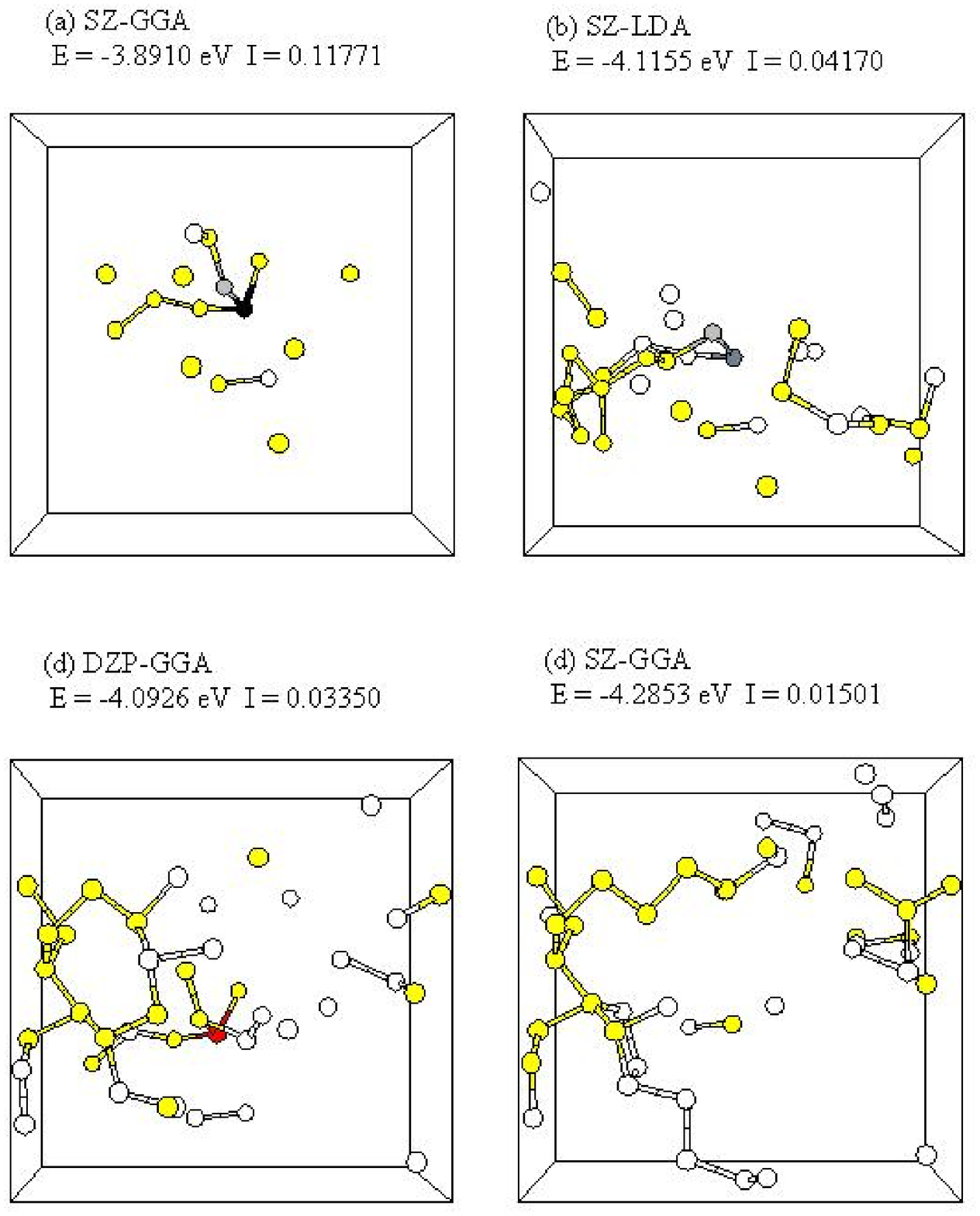}
\caption{\label{space-far-rel}
(Color Online) Spatial character of localized eigenstates for the HOMO 
state for the relaxed FAR model. The color coding is the 
same as the CLOSE case in Fig.~\ref{space-close}.
}
\end{figure}

Finally, we visualize a chosen state (HOMO in this case)
for the relaxed FAR model using a color 
coding in Fig.~\ref{space-far-rel}
The Mulliken charge concentrations on the atoms changes from confined
cluster-like character (or equivalently short 
1D strings) in Fig.~\ref{space-far-rel}(a)
to long string-like
character of atoms in Fig.~\ref{space-far-rel}(d) as 
one tunes the basis and functional from
SZ-GGA through to DZP-LDA . The "tinker-toy" character can be
attributed to less localized states and it is mainly due to
weak quantum mechanical mixing.
This behavior has been observed earlier by 
Drabold {\it et. al}.~\cite{Dong,Nak}


\section{CONCLUSION}

We have performed a first principles electronic structure calculation on 
three Si supercells: two 216-atom supercells of amorphous silicon with 
two dangling bonds and one 218-atom supercell of hydrogenated crystalline 
silicon with a void. Depending on the initial 
distance between the dangling bonds, 
the two a-Si models have been classified as CLOSE and FAR. We examined 
the nature of localized band tail and gaps states within the LDA and GGA 
using both minimal SZ basis and more complete DZP basis with particular 
attention paid to relaxation effects. 
Spin localization and wave function 
localization for dangling bond defect states 
states has also been studied. We computed the wave function localization as 
the spread of the wave function in real space and via other measures that 
utilize the Mulliken charges.

For the frozen lattice calculations, we find 
that the localization of wave functions 
associated with defect states 
decrease with larger basis sets and has enhancement of localization using GGA 
compared to LDA for all the models. 
The reduction in charges at  
the atomic sites for a larger basis set can be attributed to the  
hybridization between the atomic orbitals, providing the electronic 
charges more degrees of freedom to redistribute themselves. This
is reflected in a smaller distance between the defect
states and the band gap edges, that also favors delocalization.
Unpaired spin LSDA calculations performed on 
frozen lattices showed that the degrees of spin and wave
function localization
are different. In particular, degree of spin 
localization at a dangling bond site is
far greater than the degree of wave function localization.
The difference between the localization 
of a defect state in a fully relaxed and frozen 
systems is non-trivial, especially in 
the minimal basis calculations.
In particular, 
there was a considerable reduction in localization 
(as measured using the Mulliken charge) 
for the relaxed systems compared to the 
frozen lattices. We also conclude that a large 
basis set (DZP in this case) is necessary for an accurate description of
both the geometry and localized states associated with defect sites. 

\begin{acknowledgments}

We are grateful to Professor Normand Mousseau 
for sending us his models of amorphous 
silicon. We also acknowledge the support of 
the National Science Foundation under 
Grant Nos. DMR-0074624, DMR-0310933 and DMR-0205858.
P.O. acknowledges support for his research visit
to Ohio University from the Programa de Movilidad
de Investigadores of Ministerio de Educaci\'on y Cultura
of Spain.

\end{acknowledgments}

\newpage

\begin{center}
\begin{table}[htbp]
\caption{\label{tab1}The frozen lattice results for the spread 
$\sigma^2$ and the charge integrated over a sphere $q_\Omega$
for the localized midgap state M for the supercell \emph{c}-Si:H. 
Unit for $\sigma^2$ is \AA$^2$.}
\begin{ruledtabular}
\begin{tabular}{lllcc}
Functional & Basis & $\sigma^2$ & $q_\Omega$\\
\hline
LDA & DZP & 32.14 & 0.80 \\
GGA & DZP & 28.46 & 0.83\\
LDA & SZ  & 19.44 & 0.90 \\
GGA & SZ  & 18.55 & 0.91 \\
\end{tabular}
\end{ruledtabular}
\end{table}
\end{center}

\begin{center}
\begin{table}[htbp]
\caption{\label{tab2}The spread 
and integrated charge for 
the localized midgap state M for the relaxed \emph{c}-Si:H model.}
\begin{ruledtabular}
\begin{tabular}{lllcc}
Functional & Basis & $\sigma^2$ & $q_\Omega$\\
\hline  
LDA & DZP & 34.24 & 0.79 \\
GGA & DZP & 27.14 & 0.84 \\
LDA & SZ  & 38.42 & 0.76 \\
GGA & SZ  & 32.07 & 0.81 \\
\end{tabular}
\end{ruledtabular}
\end{table}
\end{center}

\begin{center}
\begin{table}[htbp]
\caption{\label{tab3}The results for $\sigma^2$  and $q_\Omega$
corresponding to the localized LUMO state for the relaxed
CLOSE model.}
\begin{ruledtabular}
\begin{tabular}{lllcc}
Functional & Basis & $\sigma^2$ & $q_\Omega$\\
\hline        
LDA & DZP & 31.85 & 0.76 \\
GGA & DZP & 28.14 & 0.79\\
LDA & SZ  & 23.50 & 0.84 \\
GGA & SZ  & 22.45 & 0.87 \\
\end{tabular}
\end{ruledtabular}    
\end{table}
\end{center}


\begin{thebibliography}{99}
\bibitem{ed} The information entropy 
functional is the only form which consistently
represents our state of knowledge in a probability 
theoretic sense. For a discussion of the 
$p log(p)$ functional versus $p^2$ (as for the 
IPR of this paper), see Jaynes (Ref. \onlinecite{Jaynes}) 
and references therein.
\end{thebibliography}
\end{document}